# A 3D pancreatic cancer model with integrated optical sensors for non-invasive metabolism monitoring and drug screening

Anna Chiara Siciliano[a,b,1], Stefania Forciniti[a,1], Valentina Onesto[a], Helena Iuele[a], Donatella Delle Cave[c], Federica Carnevali[a,b], Giuseppe Gigli[a,d], Enza Lonardo[c] and Loretta L. del Mercato[a,*]

[a] Institute of Nanotechnology, National Research Council (Cnr-NANOTEC), c/o Campus Ecotekne, via Monteroni, 73100, Lecce, Italy.

[b] Department of Mathematics and Physics ''Ennio De Giorgi", University of Salento, c/o Campus Ecotekne, via Monteroni, 73100, Lecce, Italy

[c] Institute of Genetics and Biophysics Adriano Buzzati-Traverso, National Research Council (Cnr-IGB), Naples, 80131, Italy.

[d] Department of Experimental Medicine, University of Salento, c/o Campus Ecotekne, via Monteroni, 73100, Lecce, Italy

[1] Equal contribution
[*] Corresponding author: loretta.delmercato@nanotec.cnr.it (+39 0832-3-19825)

**Abstract**
A distinct feature of PDAC is a prominent tumor microenvironment (TME) with remarkable cellular and spatial heterogeneity that meaningfully impacts disease biology and treatment resistance.
The dynamic crosstalk between cancer cells and the dense stromal compartment leads to spatially and temporally heterogeneous metabolic alterations, such as acidic pH that contributes to drug resistance in PDAC. Thus, monitoring the extracellular pH metabolic fluctuations within the TME is crucial to predict and to quantify anti-cancer drug efficacy. Here, we present a simple and reliable alginate-based 3D PDAC model embedding ratiometric optical pH sensors and co-cultures of tumor (AsPC-1) and stromal (PSCs) cells for simultaneously monitoring metabolic pH variations and quantify drug response. By means of time-lapse confocal laser scanning microscopy coupled with a fully automated computational analysis, the extracellular pH metabolic variations were monitored and quantified over time during drug testing with gemcitabine, folfirinox and paclitaxel, commonly used in PDAC therapy. In particular, the extracellular acidification is more pronounced after drugs treatment, resulting in increased anti-tumor effect correlated with apoptotic cell death. These findings highlight the importance of studying the influence of cellular metabolic mechanisms on tumor response to therapy in 3D tumor models, this being crucial for the development of personalized medicine approaches.

**1. Introduction**

Pancreatic ductal adenocarcinoma (PDAC) is one of the most malignant and aggressive disease with the highest mortality rate among all relevant cancers [1–3]. Nowadays, PDAC is the fourth most frequent cause of cancer-related deaths with a 5-year survival rate of less than 7% and is expected to become the second leading cause of global cancer mortality by 2030 [4]. Surgical resection followed by chemotherapy represents the best treatment option for PDAC [5]. Unfortunately, only 15-20% of patients present a potentially resectable tumor at the time of diagnosis, while the majority of them are diagnosed with late-stage disease [4,6]. This is due to the lack of early detection methods as well as the absence of obvious symptoms appearing when the tumor has progressed and metastasized to other sites. Chemotherapy remains the preferred treatment, especially in cases of advanced pancreatic tumors, even if it achieves only limited results in improving patients' survival, mainly due to chemo-resistance [7]. Multiple chemotherapy regimens have been approved for first-line treatment of PDAC, such as gemcitabine, a deoxycytidine nucleoside analog. The administration of this chemotherapeutic agent has provided only modest improvements in survival advantage to patients, showing a low rate of therapeutic efficacy [8]. However, gemcitabine is still employed in clinical use as monotherapy or in combination with paclitaxel, a taxane derivative that acts by blocking cell cycle and causing apoptosis [9,10]. Among other classes of chemotherapeutics, the multi-drug regimen folfirinox (5-fluorouracil, irinotecan and oxaliplatin) significantly improves response rates and disease-free survival in metastatic PDAC patients, though folfirinox is associated with highly toxic and adverse effects [11,12]. Several studies have demonstrated that the main cause of the current chemotherapy failure is the heterogeneous stroma-rich tumor microenvironment (TME), typical of PDAC [13]. Notably, pancreatic TME is highly dynamic with spatial and temporal heterogeneity in composition, due to genetic alterations in cancer cells that adapt the neighboring environment into a tumor-supportive niche. The continuous crosstalk between tumor cells and the surrounding extracellular matrix as well as other type of cells such as cancer associated fibroblasts (CAFs), pancreatic stellate cells and immune cells, promote intra-tumor heterogeneity and influence the response to chemotherapy treatments [14,15]. In addition, the establishment of selective pressures within the TME, such as hypoxia or acidosis derived from the altered perfusion and metabolic preferences of the different cell populations contribute not only to tumor initiation and progression but also to the poor efficacy of therapeutic drugs [16–19]. In this context, the extracellular acidification in the TME is known as a major hallmark of PDAC that occurs due to a metabolic reprogramming of cancer cells towards an increased aerobic glycolysis producing cytosol-acidifying lactate (referred to as Warburg effect) [20,21]. On the other hand, the extracellular acidification is also due to the activation of different ion transporters such as $Na^+/H^+$ exchanger (NHE-1), $Na^+$-dependent and independent $HCO_3^-/Cl^-$ exchangers and the mono-carboxylate transporter (MCT), resulting in an increased release of $H^+$ ions into the extracellular space [22–24]. Therefore, monitoring extracellular pH metabolic variations, especially during chemotherapeutic treatment, is a promising strategy in order to understand how drug sensitivity is influenced by the interplay between the different components of the microenvironment and predict the efficacy of anti-cancer drugs. Nowadays, different systems are available for measuring extracellular acidification in real-time, such as standard flux analyzers and pH nanoprobes [25,26]. Despite these methods offer high pH sensitivity and other various advantages [27], they present a limited spatio-temporal resolution, provide average values of extracellular acidification as read-out, and are more invasive than optical approaches [28]. Other techniques like surface-enhanced Raman spectroscopy (SERS) are now explored for pH sensing, particularly in the context of designing dual-signal optical sensor particles for intracellular and extracellular pH imaging [29,30]. Although SERS enables high sensitivity and spatial resolution, it needs to be coupled with plasmonic nanostructures of precise and regular geometry and pH-sensing Raman reporters with specific properties [31,32]. Among the optical methods, ratiometric fluorescence-based pH microparticles have demonstrated significant advantages for real-time analysis in living cells, because of their non-invasive features and reliable, fast, and accurate measurements with high spatial and temporal resolution [33]. By employing a pH indicator dye and a non-sensitive dye as a reference, their emission intensity ratio is translated into a specific pH value by applying automated mathematical algorithms [34]. In this study, ratiometric fluorescent pH sensors previously obtained [35] are employed by using fluorescein isothiocyanate (FITC), as pH-sensitive dye, and rhodamine B isothiocyanate (RBITC), as reference dye. Growing evidence indicates that the effect of the anti-cancer drug is not only influenced by the establishment of pH gradients within the tumor tissue, but also by the application of different cell culture conditions *in vitro* [36,37]. Indeed, the success of the therapeutic treatments tested in *in vitro* systems is often different compared to the clinical outcome [38]. In this regard, *in vitro* three-dimensional (3D) cell culture platforms are proven to be relevant preclinical models compared to 2D monolayers, as they more accurately recapitulate the architecture and features of the tumor microenvironment [39,40]. Therefore, *in vitro* 3D models are essential for studying pH fluctuations in tumor and how these metabolic changes affect the efficacy of anti-cancer drugs. To this aim, 3D alginate-based microgels embedding human pancreatic tumor cells (AsPC-1), pancreatic stellate cells (PSCs), as stromal component, and optical ratiometric pH sensors were produced. Notably, alginate was chosen as natural polymer for producing the spherical hydrogels because of its high biocompatibility, its ability to allow diffusion of nutrients and chemotherapeutic drugs [41], and its transparency that makes it an excellent candidate for optical microscopy [35]. Then, 3D tumor-stroma microgels were exposed to the most commonly used therapeutic regimens in PDAC treatment, such as folfirinox, gemcitabine and paclitaxel. Following drug administration, the extracellular pH metabolic variations were monitored over time and space through CLSM live cell imaging. Interestingly, the influence of the extracellular pH variations upon drug efficacy and acquired chemo-resistance was correlated, taking into account the dynamic crosstalk existing within pancreatic tumor and stromal cells.

2. **Matherials and methods**

*2.1 Chemicals*
Rhodamine B isothiocyanate (RBITC), Fluorescein 5(6)-isothiocyanate (FITC), Calcein AM, propidium iodide, Alginic acid sodium salt from brown algae, Calcium chloride dihydrate, Irinotecan hydrochloride (IRI), Oxaliplatin (OX) and 5-Fluorouracil (5-FU) were all purchased from Sigma-Aldrich, Darmstadt, Germany. Gemcitabine HCl was purchased from Selleckchem, Paclitaxel and Cell Mask$^{TM}$ Deep Red Plasma Membrane Stain from Invitrogen, Thermo fisher Scientific. Tetraethyl orthosilicate (TEOS) and (3-Aminopropyl) triethoxysilane (APTES) were purchased from Aldrich chemistry. Anhydrous ethanol from VWR, ethanol from Honeywell, Potassium Chloride (KCl) from Sigma life science.

*2.2 Synthesis and characterization of SiO$_2$-based ratiometric optical pH sensors*

Synthesis and characterization of silica ($SiO_2$) pH sensors were obtained as previously reported [35]. Briefly, $SiO_2$ seed particles were synthesized adopting a modified Stöber method [42]. The obtained $SiO_2$ microparticles were further functionalized with FITC-APTES and RBITC-APTES thiourea and purified by means of serial centrifugations. The morphology of the pH sensors was analyzed by scanning electron microscopy (SEM, Sigma 300VP, Zeiss, Germany), using 3 kV as accelerating voltage and a secondary electron detector (SE2). Sensors were deposited onto a silicon wafer and sputter coated with a 10 nm thick gold layer (Safematic CCU-010 LV Vacuum Coating) prior to their observation. Acquired images were then analyzed in ImageJ [43] software to extract sensor diameters and distributions. Hydrodinamic diameter, monodispersity and surface charge of the pH sensors dispersed in D.I. water (refractive index 1.458, absorption 0.010, 25°C, 3 min equilibrium time [44]) were measured by Dynamic Light Scattering (DLS) adopting a Zetasizer Nano ZS purchased from MALVERN and data analyzed Zetasizer 7.12 software provided by the manufacturer setting. The $SiO_2$-based pH sensors were then characterized through a fluorometer (ClarioStar BMG Labtech, Germany) by evaluating their response in the range of physiological pHs (range 5.0, 6.0, 7.0, 8.0) and their reversibility at day 0 and after 7 days of aging through series of three switches between pH 7.0 and pH 5.0. pH sensors were also calibrated in L-15-adjusted media in the range of physiological pHs (5.0, 6.0, 7.0, 8.0) by Confocal Laser Scanning Microscopy (CLSM, Zeiss LSM 700, Germany) (laser line 488 nm for excitation and 500–550 nm for emission of FITC; laser line 555 nm for excitation and 570–620 nm for emission of RBITC).

### *2.3 Cell lines*

Human pancreatic cancer cell line AsPC-1 (ATCC CRL-1682™) was cultured at 37 °C in a humidified 5% $CO_2$ incubator and grown in RPMI-1640 (Sigma-Merck KGaA, Darmstadt, Germany) supplemented with 10% FBS (Gibco), 2 mM glutamine and 1% penicillin/streptomycin (Sigma-Merck KGaA, Darmstadt, Germany). Human immortalized pancreatic stellate cells (PSCs), kindly provided by Dr. Enza Lonardo, were cultured in DMEM medium (Sigma-Merck KGaA, Darmstadt, Germany) supplemented with 10% FBS (Gibco), 2 mM glutamine and 1% penicillin/ streptomycin (Sigma-Merck KGaA, Darmstadt, Germany) at 37°C with 5% $CO_2$.

### *2.4 Generation of 3D tumor-stroma microgels*

The 3D tumor-stroma microgels were fabricated by employing a lab-made microencapsulation system, as previously described [35]. Particularly, AsPC-1 tumor cells, pre-labelled with CellTracker™ Deep Red (Invitrogen, ThermoFisher Scientific), and PSCs were mixed in a 1:3 ratio ($4 \times 10^6$ of total cells) in 250 μL of fresh medium. Then, 250 μL of alginate solution (3% in distilled water) were added to the cell suspension with 40 μL of pH sensors ($5.38 \times 10^6$ particles/mL of the pH sensors stock solution), previously stirred for 5 minutes. Finally, the resulting mix was loaded into a syringe (BD Plastipak™ 1-mL Syringe) and placed into a syringe pump (World Precision Instruments, Model AL-4000). Subsequently, a high potential difference was applied by connecting the cathode of the generator to the needle of the syringe (21 G blunt needle), while the anode was immersed in a Petri dish containing 12 mL of $CaCl_2$ solution (0.1 M $CaCl_2$ and 0.4% w/v of Tween-20 dissolved in deionized water) positioned under the syringe tip at a distance of 3 cm. Then, the syringe pump was set up at a constant flow rate of 0.05 mL min$^{-1}$ and by applying high voltage (V 4.5), alginate microgels were produced once the drops fell into the calcium chloride solution. The obtained tumor-stroma alginate microgels were washed three times with fresh culture medium in order to remove the $CaCl_2$ solution and maintained in the incubator, before being acquired at the CLSM. In parallel, cell-free alginate microgels embedding only pH sensors were produced, in order to test sensors photobleaching in the alginate scaffolds.

## 2.5 Determination of $IC_{50}$

AsPC-1 or PSC cells were counted by Trypan Blue dye exclusion and $5 \times 10^3$ cells were seeded in 96-well cell culture plates. A similar procedure was performed for seeding AsPC-1 and PSC co-culture in a 1:3 ratio, respectively. After 24 hours, cell lines were treated with a range of concentrations from 0 to 100 µM of paclitaxel, folfirinox (oxaliplatin, 5-fluorouracil, irinotecan) and gemcitabine. Then, cell viability was measured by PrestoBlue™ Cell Viability Reagent (Invitrogen, ThermoFisher Scientific), following the instructions of the manufacturer's protocol. 5 µL of PrestoBlue solution were added directly in the medium of each well and after 1 hour of incubation at 37°C, the fluorescent signal was measured by using a plate reader (CLARIOstarplus BMG LABTECH), at 535 nm of excitation wavelength and 590 nm of emission wavelength. Data obtained for each treatment were normalized with the respective control group. Finally, the half maximal inhibitory concentration ($IC_{50}$) values were extrapolated from the sigmoidal curves obtained by plotting the cell proliferation rate against the various drug concentrations on a semi-logarithmic scale in GraphPad Prism version 8.0 software.

## 2.6 Live/dead assay

Cell viability of the 3D alginate microgels encapsulating PSC and AsPC-1 cells was first evaluated after 10, 24, and 48 hours of treatment with each chemotherapeutic drug and 10 hours after a second drug administration (58 hours). Particularly, after being produced, the 3D tumor-stroma microgels were treated with paclitaxel (8.9 µM), folfirinox (4.7 µM 5-fluorouracil, 37 µM oxaliplatin, 6.2 µM irinotecan) and gemcitabine (9.3 µM) and at each time point 0.25 µM of Calcein AM (Sigma-Aldrich, Darmstadt, Germany) and 10 µM of propidium iodide (PI, Sigma-Aldrich, Darmstadt, Germany) were added to each sample for staining live and dead cells, respectively. The untreated microgels were used as control. After 30 min of incubation, the alginate microgels were washed with complete medium and placed in an 8-well chamber slides (IBIDI, Berlin, Germany), previously functionalized with 0.2 mg/ml of poly-L-lysine (Sigma-Aldrich, Darmstadt, Germany) in order to avoid the hydrogels shift during CLSM acquisition. Representative images were acquired at 10, 24, 48, and 58 hours of culture on untreated and treated alginate microgels, by using a CLSM (Zeiss LSM700, Germany) at 20X magnification. The maximum projections of z-stack images and the semi-automatic particle analysis was performed by Image J software in order to estimate the percentage of live cells in 5 hydrogels for each condition.

## 2.7 Annexin V staining for apoptosis evaluation

Apoptotic cell death was detected using FITC Annexin V (Biolegend, San Diego), according to the manufacturer's recommended protocol. Briefly, the 3D alginate microgels encapsulating PSC and AsPC-1 cells were treated with paclitaxel (8.9 µM), folfirinox (4.7 µM 5-fluorouracil, 37 µM oxaliplatin, 6.2 µM irinotecan), and gemcitabine (9.3 µM) for 10, 24 and 48 hours. The untreated alginate microgels were used as control. At each time point, tumor-stroma microgels were incubated with EDTA 0.5 M in order to dissolve alginate instantly and extract cells from them. The samples were washed twice in PBS and $2 \times 10^5$ cells were re-suspended in 20 µl of Annexin V binding buffer. Then, cells were stained with 1µl of FITC-AnnexinV and Propidium iodide (PI) (20 µg/mL) for 15 min at room temperature. Then, 280 µl of Annexin V binding buffer were added in each sample just before the analysis with the CytoFLEX S flow cytometer (Beckman Coulter, USA). Approximately 20.000 events were acquired for each sample and analyzed by CytExpert software. Dead cells, debris, and doublets were excluded based upon forward scatter and side scatter measurements. Viable cells were defined as annexinV-negative and PI-negative; early apoptotic cells were defined as annexinV-positive and PI-negative; late apoptotic/necrotic cells were defined as annexin V-positive and PI-positive.

## *2.8 Total RNA extraction and Real-Time quantitative PCR*

The 3D tumor-stroma microgels were produced and treated with the chemotherapeutic drugs as described above (section 2.3 and 2.6). After treatment, cells were extracted from alginate microgels by incubating them with EDTA 0.5 M. Then, cells were washed twice in PBS 1X and pellets were collected by centrifugation at 1000 x g for 10 min at 4°C for RNA extraction. Total RNA was extracted from each pellet using the Total RNA Purification Plus Kit (Norgen Biotek Corp., Canada). One microgram of total RNA was used for cDNA synthesis with High-Capacity reverse transcriptase (Thermofisher). Then, quantitative real-time PCR was performed using a SYBR Green PCR master mix (Thermofisher), according to the manufacturer's instructions. The list of utilized primers is depicted in **Table 1**.

**Table 1**. Primers used for Real-Time quantitative PCR.

| Gene symbol | Forward primer (5'->3') | Reverse primer (5'->3') |
|---|---|---|
| *GAPDH* | CAGGAGCGAGATCCCT | GGTGCTAAGCAGTTGGT |
| *ABCB1* | TGACATTTATTCAAAGTTAAAAGCA | TAGACACTTTATGCAAACATTTCAA |
| *ABCG1* | TCAGGGACCTTTCCTATTCG | TTCCTTTCAGGAGGGTCTTGT |
| *ABCG2* | TCATGTTAAGGATTGAAGCCAAAGGC | TGTGAGATTGACCAACAGACCTGA |

## *2.9 Live cell imaging*

The calibration of ratiometric optical pH sensors embedded in 3D alginate microgels without cells was performed by CLSM (Zeiss LSM 700, Carl Zeiss AG). To this aim, the samples were placed in an 8-well chamber slides (IBIDI, Berlin, Germany), previously functionalized with 0.2 mg/ml of poly-L-lysine. Then, alginate microgels were exposed to different pH-adjusted cell media (pHs 7.0, 6.0, 5.0, 4.0) and they were left to stabilize for 10 minutes before being acquired, along the z-axis, via CLSM equipped with Okolab Stage Top Incubator (Okolab s.r.l., Italy) for controlled $CO_2$ and temperature. After the calibration measurements, 3D tumor-stroma microgels were incubated in the presence or absence of each chemotherapeutic drug and placed into the Okolab chamber. The pH variations were first monitored in time-lapse for 10 hours, and next at different time-points (10, 24, 48, and 58 hours). For time-lapse and end-point acquisitions, z-stack images of the whole microgels were acquired at interval of 1 hour; FITC was excited by using the 488 nm laser line (0.8%), RBITC by the 555 nm laser line (0.6%) and the cancer cells stained with Deep Red were excited by using the 639 nm laser line (1%).

*2.10 4D image processing and analysis*

4D (x,y,z,t) CLSM images were automatically analyzed in GNU Octave (version 8.2.0) with a custom algorithm [35,45] developed to extract pH values measured by ratiometric optical sensors embedded in 3D structures. Briefly, input of the algorithm were the fluorescence indicator ($Im_{FITC}$) and reference ($Im_{RBITC}$) channel images of the sensors. Each z-slice of $Im_{RBITC}$ was separately converted to grayscale, median filtered to reduce photon shot noise, binarized with Otsu's method [46] and finally segmented by a watershed transformation [47]. Then, the pH microparticles were reconstructed in 3D by direct connectivity of the binary RBITC images along the z-axis, resulting in a 3D binary matrix, which was used as a mask to extract positions and mean pixel-by-pixel ratio ($I_{FITC}/I_{RBITC}$) of the fluorescence intensities of the sensors belonging to the original $Im_{FITC}$ and $Im_{RBITC}$ images. Finally, $I_{FITC}/I_{RBITC}$ values were passed in a previously extracted calibration curve to obtain the pH read-outs. The code was repeated iteratively for each time point $t$ and pH was also monitored globally over time by calculating, for each $t$, mean and standard deviation of the pH measured by the single sensors.

*2.11 Statistical analysis*

Experiments were performed in triplicate and the error bars refer to either standard deviation (SD) or standard error of the mean (SEM). One-way analysis of variance (ANOVA) was performed to compare multiple conditions, and the Student's t-test was used for individual group comparison. Differences with p-values < 0.05 were considered significant.

## 3. Results and discussion

*3.1 Properties of ratiometric optical pH sensors embedded within 3D tumor-stroma microgels.*

To reproduce 3D *in vitro* tumor-stroma models for non-invasive *in situ* monitoring of cell metabolism and drug screening, silica-based fluorescent pH sensors based on fluorescein isothiocyanate (FITC) as pH-reference dye and rhodamine B isothiocyanate (RBITC) as pH-insensitive dye were synthetized following the protocol described by Rizzo et al. (2022) [35] and included into alginate microgels. The morphology of the microsensors was evaluated by scanning electron microscopy (SEM) (**Fig. S1a**), evidencing uniform, spherical, smooth and monodispersed microparticles with a mean diameter of 1101 ± 43 nm (**Fig. S1b**), moreover dynamic light scattering analysis (DLS) reveled a hydrodynamic diameter of 2307 ± 67,57 nm (Pdl 0,192) and a negative superficial charge of -77,1 ± 0,71 mV (**Fig. S1c-d**).

Next, the possible cytotoxic effect of the embedded fluorescent pH sensors on alginate microgels encapsulating AsPC-1 tumor cells and PSC stromal cells in a 1:3 ratio, respectively was evaluated by calcein AM/propidium iodide live/dead assay. CLSM images reported in **Figure S2a,** show that in the presence of the fluorescent pH sensors, cells remained viable (green) until day 6, with the occurrence of only few red spots indicative of dead cells. In parallel, the histogram reporting the quantification of the number of live cells at days 1, 3 and 6 **(Fig. S2b)** demonstrates that the percentage of live cells increased over time, indicating that no cytotoxic effects of the embedded SiO$_2$-based pH sensors was reported.

Next, the response of pH sensors to different pHs was evaluated within the alginate-based microgels. First, we confirmed the pH sensors sensitivity **(Fig. S3a, b)**, and the photostability of both the FITC and the RBITC dyes **(Fig. S3c, d)**, embedded into the alginate microgels without cells, thus demonstrating a stable trend also maintained in the intensity ratio ($I_{FITC}/I_{RBITC}$) curve **(Fig. S3e)** for up to 18 hours of irradiation under the CLSM, guaranteeing therefore good reliability of the pH measurements during the 10 hours of our timelapse experiments. Then, the calibration of the whole system was performed. **Figure 1a** shows representative CLSM micrographs of 3D tumor-stroma microgels exposed to different pH-adjusted cell media (range 4.0-7.0) and acquired, along the z-axis, under controlled temperature (37 °C) and 5% $CO_2$. As expected, the fluorescence intensity of the pH-sensitive dye (FITC), increases with high pH values, while the fluorescence intensity of the reference dye (RBITC) remains unchanged, thus demonstrating the sensitivity of $SiO_2$ microparticles to pH variations. Therefore, plotting the intensity ratio ($I_{FITC}/I_{RBITC}$) as a function of pH, a calibration curve was obtained **(Fig. 1b)** indicating a linear correlation between the fluorescence intensity and the analyte concentrations ($H^+$ ions). The same pH values recorded during CLSM acquisition were plotted as 3D colorimetric maps **(Fig. 1c)**, whose colors replicate the one shown by the FITC/RBITC overlay at each pH, from red (pH 4.0) to green (pH 7.0). This procedure was performed at the beginning of each extracellular pH monitoring experiment.

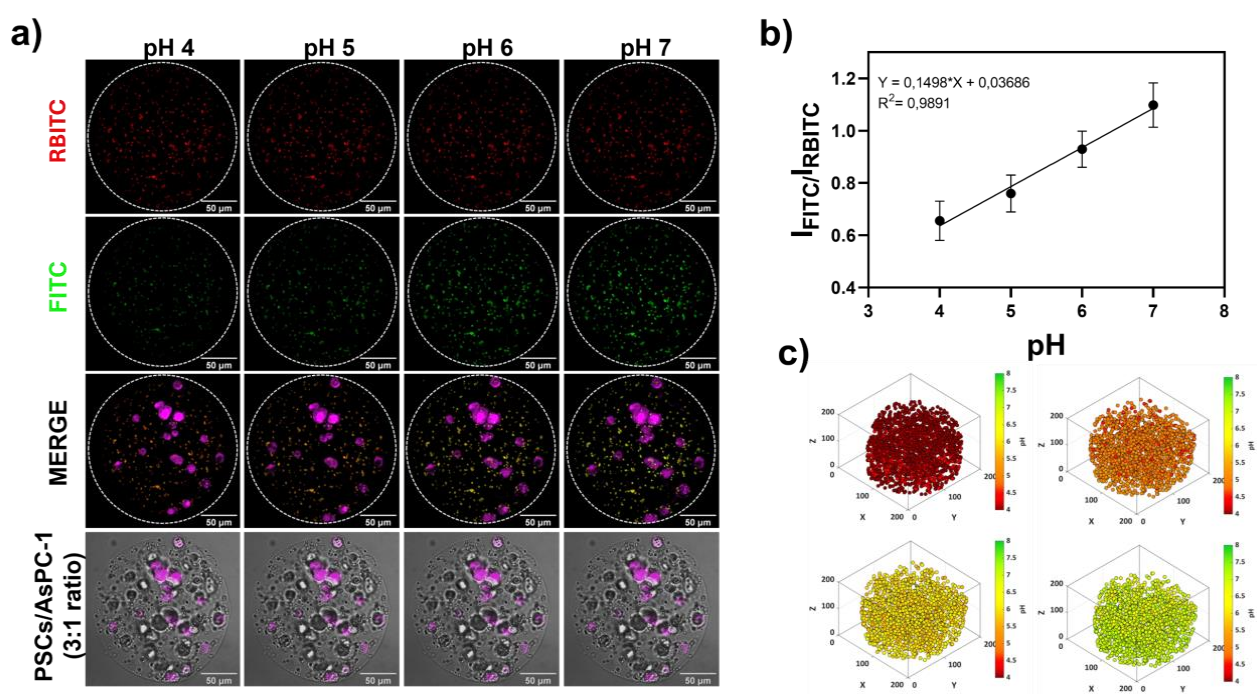

**Figure 1. Calibration of ratiometric optical pH sensors in 3D tumor-stroma microgels. a)** Representative CLSM images of alginate microgels embedding pH sensors, AsPC-1, and PSC cells exposed to different pH-adjusted cell media (pHs 4.0, 5.0, 6.0, 7.0). RBITC (red channel), FITC (green channel), PSCs in bright field (BF, grey channel), and AsPC-1 (Deep Red, magenta channel) are reported. Maximum z-projections were extrapolated from 35 z-stacks for each pH. Z-stack step size = 2.55 μm. Scale bars, 50 μm. **b)** Calibration curve of pH sensors reporting the fluorescence intensity ratio of FITC (green channel) and RBITC (red channel) as function of pH. Data are means ± SEM. **c)** 3D colormaps of spatial distribution of pH sensors within the alginate microgels. The fluorescence intensity ratio of each sensor was extrapolated and converted in false color for each pH value.

*3.2 Non-invasive tumor-stroma extracellular pH mapping during drug testing*

Notably, PDAC is characterized by high chemoresistance against the standard therapy. These phenomena are mainly due to the metabolic aberrations of pancreatic cancer cells, first of all an increased glucose consumption [48]. In fact, as happens in most solid tumors, also PDAC cells reprogram their metabolism towards an increased aerobic glycolysis and a reduced mitochondrial oxidative phosphorylation, referred to as "Warburg effect" [49–51]. This metabolic switch results in a higher production and then secretion of lactate and protons from glycolytic cells, leading to an extremely acidic and tumor-supportive extracellular microenvironment [51–53]. The acidic pH affects cancer cell behaviors, including epithelial to mesenchymal transition (EMT) [54,55], uncontrolled proliferation, local invasion [56], and mainly chemoresistance [56,57]. Hence, monitoring the local pH variations in the extracellular tumor microenvironment during chemotherapy treatment is crucial for better understanding the biology of PDAC and the role of pH fluctuations on drug response. In this regard, 3D alginate microgels, embedding AsPC-1 tumor cells, PSC stromal cells, and FITC/RBITC pH sensors, were generated and treated with chemotherapeutic drugs used in PDAC standard therapy (paclitaxel, folfirinox and gemcitabine). In order to administer the appropriate drug concentration, the $IC_{50}$ values were previously determined for each chemotherapeutic drug. In particular, $IC_{50}$ values were extrapolated from the sigmoidal curves for AsPC-1 (**Fig. S4a**) and PSC (**Fig. S4b**) monocultures and for AsPC-1 and PSC co-cultures (**Fig. S4c**); the obtained concentrations are indicated in the table reported in **Figure S4d** and $IC_{50}$ values of co-cultures were then used for the treatment of 3D alginate tumor microgels. Hence, to monitor pH variations over time and space, 3D tumor-stroma microgels were produced (See Materials and Methods and **Fig. 2b**) and imaged for 10 hours through CLSM in time lapse mode, with controlled temperature (37°C) and 5% $CO_2$. In **Figure 2a**, representative CLSM images for both untreated and treated microgels recorded at 1 hour and 10 hours are reported, and the other time points (from 2 hours to 8 hours) are shown in **Figure S5**. The pH calibration curves were obtained before starting each time lapse imaging (data not shown). Moreover, the fluorescent intensity ratio of pH sensors was calculated and the corresponding pH values were converted in color. Each alginate microgel was represented as a 3D colorimetric map at 1 hour and 10 hours, and the topographical reconstruction of the dynamic variations of pH sensors was monitored over time and space (**Fig. 2a and Fig. S5**). The pH variations recorded over time within the 3D tumor-stroma microgels are shown in **Figure 2c**; in particular, compared to control, all drugs tested determined a general acidification of the extracellular compartment within the 3D tumor-stroma microgels. Interestingly, after 10 hours of treatment more acidic pH values were observed (pH 5.2 for paclitaxel, pH 5.8 for folfirinox, pH 5.7 for gemcitabine) than in the first hour (pH 7.5). On the contrary, the untreated microgel shows a negligible acidification reaching a maximum value of 6.9. This result is confirmed also from the 3D colorimetric maps, whose colors change according to the pH value. Taken together, these data provide a possible correlation between the observed extracellular acidification and the cell inhibition mechanisms induced by the chemotherapy treatments. In fact, it is known that all drugs tested cause apoptotic cell death [58–60], therefore accompanied by cellular morphological changes, including shrinkage and apoptotic bodies release which led to an accumulation of acidic waste in the surrounding environment causing acidification [61].

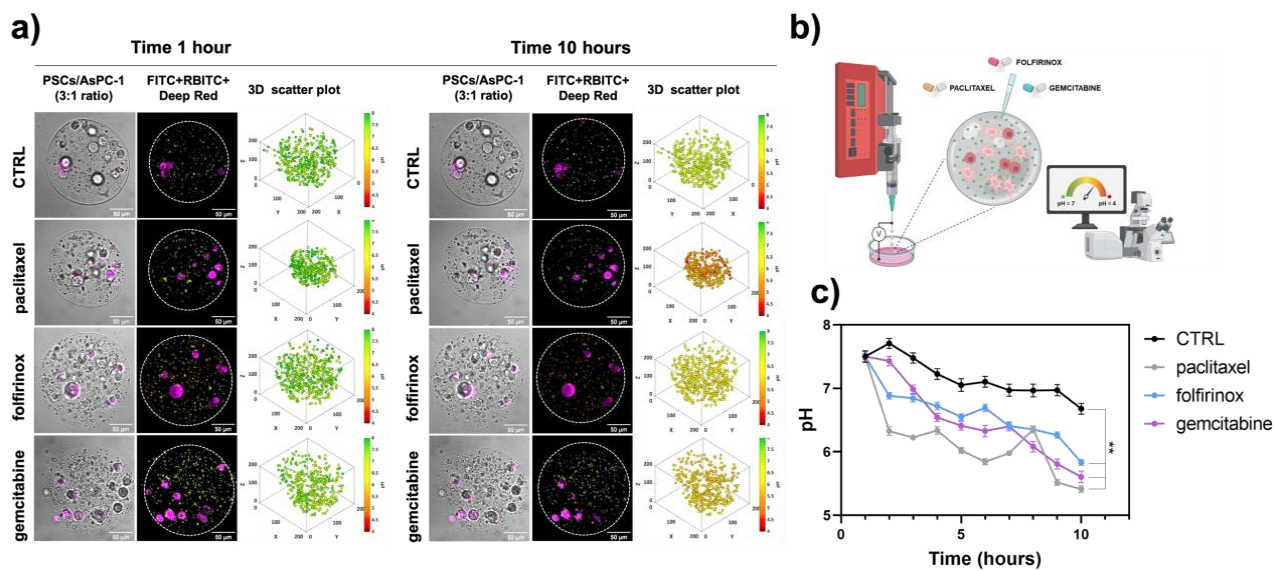

**Figure 2. Tumor-stroma extracellular pH evaluation over time and space during drug testing. a)** Representative CLSM time lapse images recorded at 1hour (left panel) and 10 hours (right panel) of a spherical alginate microgel embedding fluorescent pH sensors, PSCs cells, and AsPC-1 cells both untreated (CTRL) and treated with paclitaxel (8.9 μM), folfirinox (Oxa 37 μM, 5-FU 4.7 μM, Iri 6.2 μM) and gemcitabine (9.3 μM). Overlay of FITC, RBITC, and Deep Red (green, red, and magenta channels) represents FITC/RBITC pH sensors and AsPC-1 tumor cells, respectively; in bright field (BF, grey channel) PSC stromal cells merged with AsPC-1 (Deep Red, magenta) are shown. Z-stack step size=2.55 μm; scale bars, 50 μm. 3D scatter plots of the pH sensors shown in the CLSM micrographs. **b)** Schematic representation of the experimental set up for the generation of tumor-stroma microgels and their pH sensing analysis during drugs testing. Objects are not to scale. **c)** Quantification of the time-lapse experiments shown in a) and in **Figure S5** reporting the pH values extrapolated from the calibration curves obtained at the beginning of each experimental condition and plotted as function of time. Data are means ± SEM. Statistical analysis: ** $p<0.01$, CTRL *vs.* paclitaxel, or folfirinox, or gemcitabine.

### *3.3 Chemotherapeutic drugs induce anti-tumor effects and apoptotic cell death in 3D tumor-stroma microgels*

In order to explain whether the extracellular acidification following drug treatments is due to cell death and to confirm that pH sensors do not influence cell behavior or drug response, both untreated and treated 3D tumor-stroma microgels were first tested for viability. In particular, PSC stromal cells and AsPC-1 tumor cells were encapsulated without pH sensors in spherical alginate microgels and treated with paclitaxel (8.9 μM), gemcitabine (9.3 μM), and folfirinox (oxaliplatin 37 μM, 5-fluorouracil 4.7 μM, irinotecan 6.2 μM) for 10 hours. Then, alginate hydrogels were stained with Calcein-AM (green) and propidium iodide (PI, red) for live/dead assay and acquired by CLSM. Data reported in **Figure 3a** show that in the untreated tumor-stroma microgels, cells are viable as noted by green nuclear staining and only few cells show red fluorescence indicative of death. On the contrary, in the pharmacologically treated microgels, especially for paclitaxel, it is possible to appreciate a high number of dead cells, as noted by red nuclear staining with the occurrence of only few green spots indicative of live cells. Then, the number of live cells (green) were quantified after 10 hours of treatment with the different drugs and expressed as percentage of the total given by the sum of red and green cells. As reported in **Figure 3b,** the percentage of live cells is strongly reduced after all drug treatments, especially with paclitaxel, demonstrating an enhanced anti-tumor effect of paclitaxel than folfirinox and gemcitabine on 3D tumor-stroma microgels. These results of increased cell death after the chemotherapeutic drugs treatment correlate with the extracellular metabolic acidification measured in the PDAC alginate microgels. This event of acidification could be due to the accumulation of acidic waste products derived from cell shrinking and fragmentation into membrane-bound apoptotic bodies as an early response to chemotherapy treatments [62]. In this framework, we next evaluated the phosphatidylserine exposure, a common event in the apoptotic cell death by performing Annexin V and PI analysis through flow cytometry. Representative dot plots and histogram reported in **Figure 3c** and **d** show the percentages of apoptotic and necrotic cells analyzed after treatment with paclitaxel, folfirinox and gemcitabine for 10 hours. In particular, at this time point, all drugs tested induced an early (Annexin V-FITC$^+$/PI$^-$) and late (Annexin V-FITC$^+$/PI$^+$) apoptosis and necrosis (Annexin V-FITC$^-$/PI$^+$) that increased over time after 24 and 48 hours of treatment **(Fig. S6)** compared to untreated control in which a basal level of cell death was observed. In particular, since the percentage of Annexin V-FITC$^+$ cells don't increase at 48 hours **(Fig. S6 c, d)** compared to 24 hours **(Fig. S6 a, b)**, the percentage of late apoptosis (Annexin V-FITC$^+$/PI$^+$) increases, especially for paclitaxel. On the contrary, after gemcitabine treatment, the apoptotic level remains stable over time suggesting that apoptosis could be only an early response to this treatment and other cell death mechanisms could be involved, such as cell cycle arrest [63].

These results confirms that the increased extracellular acidification after the chemotherapeutic drugs treatment is correlated with apoptotic cell death.

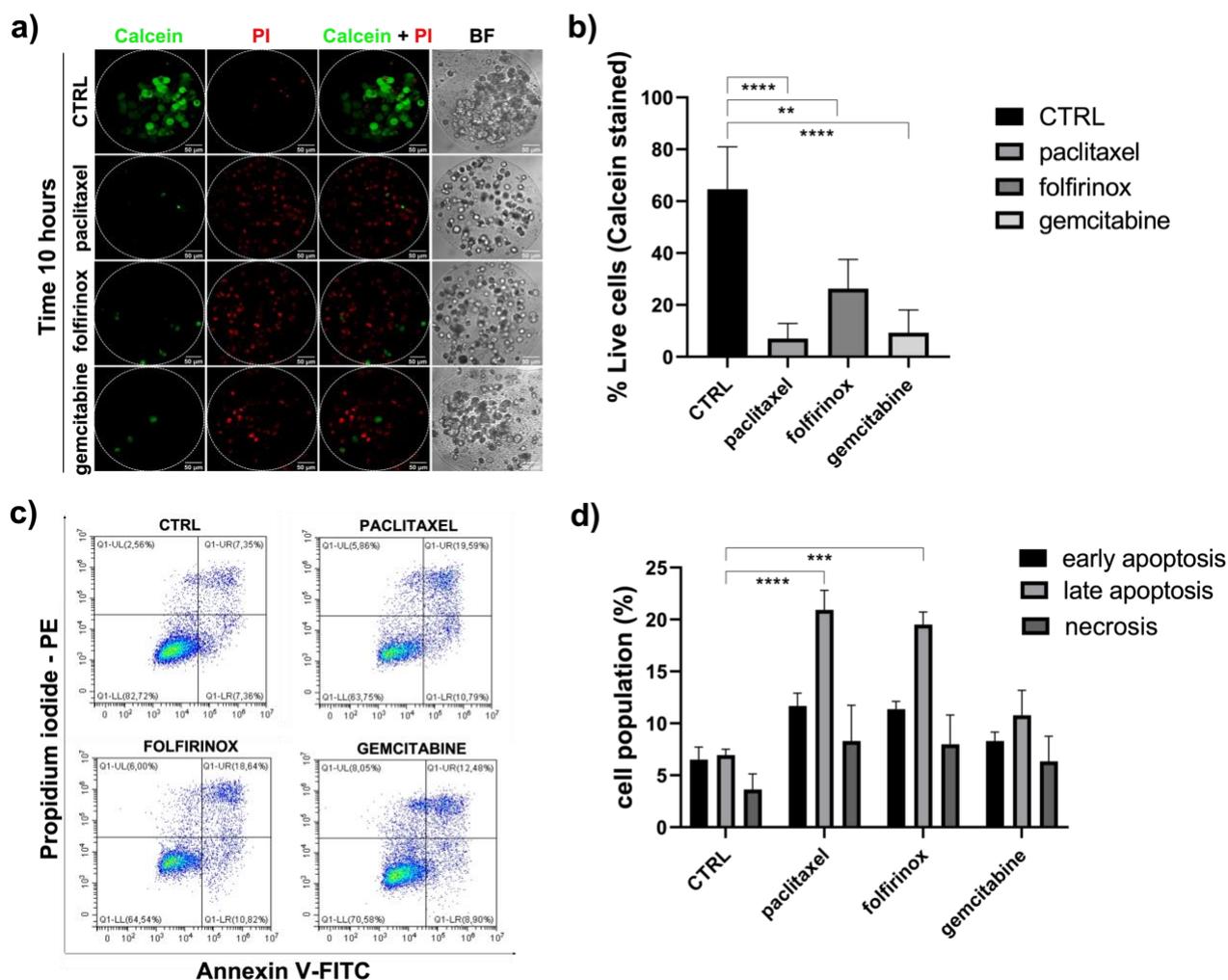

**Figure 3. Cell viability and apoptosis analysis of tumor-stroma microgels during chemotherapy treatments. a)** Representative CLSM images of live/dead-stained alginate microgels (maximum intensity projection) acquired at 10h in the presence or absence of paclitaxel (8.9 µM), gemcitabine (9.3 µM), and folfirinox (Oxa 37 µM, 5-FU 4.7 µM, Iri 6.2 µM). PSCs and AsPC-1 cells were stained with calcein AM (green channel, live cells) and PI (red channel, dead cells). In bright field (BF, grey channel) the whole alginate microgel containing the tumor/stroma cell co-culture is shown for each condition. Z-stack step size=2.55 µm; Scale bars, 50 µm. **b)** Quantification of the experiment in a) reporting the number of live cells (green) expressed as percentage of the total given by the sum of red and green cells. Data are means of 10 alginate microgels for each condition ± SEM. Statistical analysis of CTRL vs treated alginate tumor microgels * p< 0.01 and **** p< 0.0001 **c)** Representative dot plots analyzed by flow cytometry of PSC/AsPC-1 co-cultures treated with paclitaxel, folfirinox, and gemcitabine for 10 hours and stained with Annexin V-FITC (x-axis) and PI (y-axis).Q1-LR, Q1-UR and Q1-UL quadrants represent the early stage and the end stage of apoptosis/necrosis, respectively; **d)** Percentages of early apoptotic and late apoptotic/necrotic cells derived from c). Statistical analysis of CTRL vs treated alginate tumor microgels in early, late apoptosis and necrosis *** p< 0.001 and **** p< 0.0001. Data are the mean ± SEM of triplicate experiments.

## *3.4 Extracellular pH of tumor-stroma microgels correlates with long-term chemotherapy treatments*

Basing on our previous data, we reported that within 10 hours of treatment, cell death events occur in 3D tumor-stroma microgels during chemotherapeutic treatments, and consequently we observed a marked acidification of the extracellular environment. This is mainly due to the abundant release of acidic waste products from cells that undergo shrinking and fragmentation into apoptotic bodies as an early response to chemotherapy treatments [62]. However, tumor and stromal cells that did not respond to the drug activity in ten hours remain viable and may acquire drug resistance. These phenomena also happen in the clinic with PDAC patients because standard therapies do not completely eradicate the tumor which then grows back more aggressively acquiring different mechanisms of chemoresistance [64,65]. In order to evaluate a possible correlation between a prolonged chemotherapy effect and the pH trend in the extracellular tumor microenvironment, the tumor-stroma microgels were treated with paclitaxel, folfirinox, and gemcitabine for 10, 24, and 48 hours. Next, to simulate the therapeutic treatment scheme in patients [66] the alginate microgels were treated a second time with the chemotherapeutic drugs. At each time point a live/dead assay was performed for monitoring cell viability and pH fluctuations. Representative CLSM images related to these experiments are shown in **Figure 4a** and the respective quantification of live cells or pH variations after treatments are reported in **Figure 4b** and **4c**. After 10 hours of chemotherapy treatments, we observe in the treated alginate microgels a more accentuated cell death compared to the control (**Fig. 4b**), which correlates with a pH decrease (**Fig. 4c**) in the extracellular microenvironment. On the contrary, prolonging the treatment for 24 and 48 hours, a rise in pH values is noted together with an increase in the number of live cells, both in the control and in the treated microgels. This could be associated with the acquisition of drugs resistance by cells within the TME, which represents the major challenge to overcome for improving prognosis and effectiveness of treatment in PDAC patients [67,68]. Surprisingly, among the drugs tested, paclitaxel that exhibited the strongest anti-tumor effect, also causes increased cell regrowth at 24 and 48 hours, coupled with greater extracellular pH basification. This result suggests that a part of cells is resistant, grows more aggressively and duplicates more quickly. Indeed, measuring the anti-tumor effect in the 3D alginate microgels 10 hours after the second drugs administration, cell growth remains approximatively stable, indicating a greater cell chemoresistance and a reduced drug efficacy (**Fig. 4b**). On the contrary, the recorded extracellular pH again decreases reaching acidic values while in the untreated condition, the pH continues to increase reaching basic values (**Fig. 4d**) as happens 10 hours after the first drugs administration. This could be result from new apoptotic events that occur after chemotherapy treatments as previously demonstrated.

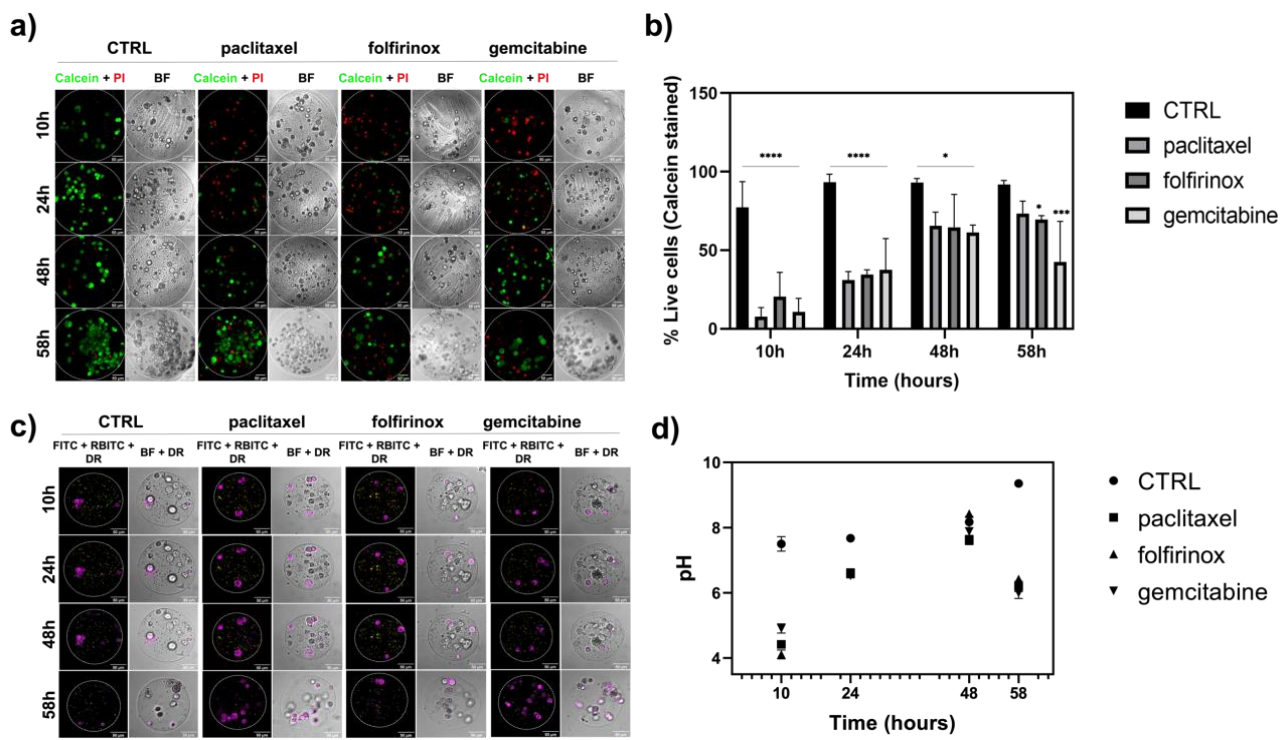

**Figure 4. Correlation between extracellular pH variations and long-term chemotherapy effect in 3D tumor-stroma microgels. a)** Representative CLSM images recorded at 10, 24, 48, and 58 hours of live/dead-stained spherical alginate microgels embedding PSCs and AsPC-1 cells both untreated (CTRL) and treated with paclitaxel (8.9 μM), folfirinox (Oxa 37 μM, 5-FU 4.7 μM, Iri 6.2 μM) and gemcitabine (9.3 μM). PSCs and AsPC-1 cells were stained with calcein AM (green channel, live cells) and PI (red channel, dead cells). In bright field (BF, grey channel) the whole alginate microgel containing tumor-stroma cell co-culture is shown for each condition. Z-stack step size=2.55 μm; scale bars, 50 μm. **b)** Quantification of the experiment in a) reporting the number of live cells (green) expressed as percentage of the total given by the sum of red and green cells. Data are means of 10 alginate microgels for each condition ± SEM. Statistical analysis of CTRL *vs* treated microgels, *p<0.05, ***p<0.001, ****p<0.0001. **c)** Representative CLSM time lapse images recorded at 10, 24, 48, and 58 hours of a spherical alginate microgel embedding pH sensors, PSCs, and AsPC-1 both untreated (CTRL) and treated with paclitaxel (8.9 μM), folfirinox (Oxa 37 μM, 5-FU 4.7 μM, Iri 6.2 μM), and gemcitabine (9.3 μM). Overlay (green, red, and magenta channels) show fluorescence signals of pH sensors and AsPC-1 tumor cells, respectively; in bright field (BF, grey channel) PSC stromal cells merged with AsPC-1 (DR, magenta) are shown. Z-stack step size=2.55 μm; scale bars, 50 μm. **d)** Quantification of the pH-sensing experiments showed in c), reporting the pH values recorded at each time-point.

### *3.5 Long-term chemotherapy treatment induces acquired resistance in 3D tumor-stroma microgels*
Chemoresistance in cancer is a significant challenge for tumor treatment [69]. In this regard, the ATP-binding cassette (ABC) transporters constitute a large family of membrane proteins which play a crucial role in the active transport of various substances across cell membranes, and consequently in multidrug resistance (MDR). Specifically, these proteins are involved in the efflux of a wide range of drugs and other xenobiotics from the inside to the outside of cells by utilizing energy derived from the hydrolysis of ATP to actively pump substances across cell membranes [70]. By doing so, they decrease the intracellular concentration of chemotherapeutic drugs, making it more challenging for these drugs to exert their cytotoxic effects on cancer cells. This efflux mechanism is a major contributor to the development of resistance to chemotherapy in cancer treatment. ABCB1, also known as P-glycoprotein (P-gp), is an ABC transporter frequently associated with MDR in cancer, whose overexpression could lead to a less sensitivity of the cells to the cytotoxic effects of the drugs, and then to chemoresistance [71]. Although ABCB1 is not directly involved in gemcitabine transport, a correlation between its expression levels and chemoresistance in PDAC has been previously observed [72]. Intriguingly, ABCB1 overexpression has been recently emerged as a novel factor in paclitaxel resistance for PDAC [73]. qPCR analyses reported in **Figure 5b** demonstrated a significant reduction in *ABCB1* expression levels 10 hours after treatment of the 3D tumor-stroma microgels with gemcitabine and folfirinox, with the exception of paclitaxel, which induces a slight increase in expression compared to the control, indicating the enhanced sensitivity of the cells to chemotherapeutic agents, in agreement with the results of cell viability and apoptosis shown in **Figure 3**. After 24, 48, and 58 hours from treatment, all the drugs induced a significant increase in *ABCB1* expression levels, indicating the acquired drug resistance of the cells, thus confirming the increased number of live cells (**Fig. 4a** and **4b**). Among ABC superfamily, also ABCG1 and ABCG2 have been identified as directly involved in PDAC chemoresistance [74]. qPCR analyses demonstrated that *ABCG1* expression is both time- and drug-dependent. Specifically, there was a decrease in *ABCG1* expression after 48 hours of treatment with all drugs, followed by an increase after 58 hours of treatment with gemcitabine and folfirinox (**Fig. S7**). Conversely, *ABCG2* levels were significantly downregulated by all drugs 24 hours after treatment, but this effect was reversed after 48 hours of treatment with paclitaxel (**Fig. 5c**) and 58 hours of treatment with gemcitabine, indicating the development of drug resistance in the cells. Altogether, these data suggest that the pH-sensing alginate microgels encapsulating tumor and stromal cells appears to be a reproducible and promising system for evaluating and monitoring *in vitro* the cellular response to drug treatment over time.

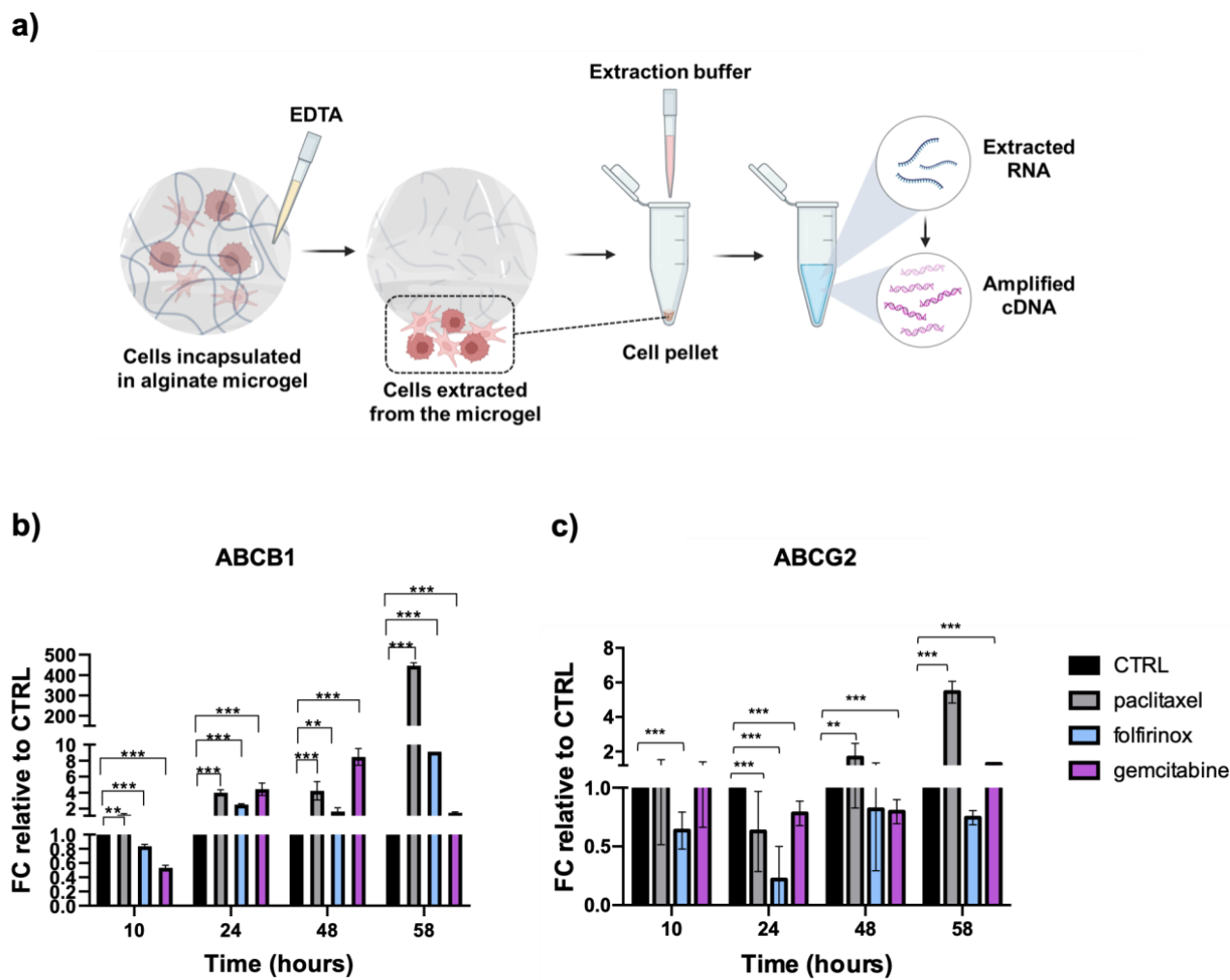

**Figure 5. Evaluation of the expression levels of markers associated with chemoresistance in 3D tumor-stroma microgels after long-term chemotherapy treatments. a)** Workflow of qPCR analysis performed on cells extracted from 3D alginate microgels. **b)** qPCR analysis of *ABCB1* gene expression of spherical alginate microgels embedding pH sensors, PSCs, and AsPC-1 both untreated (CTRL) and treated with paclitaxel (8.9 μM), folfirinox (Oxa 37 μM, 5-FU 4.7 μM, Iri 6.2 μM), and gemcitabine (9.3 μM) for 10, 24, 48, and 58 hours. **c)** qPCR analysis of *ABCG2* gene expression of spherical alginate microgels embedding pH sensors, PSCs, and AsPC-1 both untreated (CTRL) and treated with paclitaxel (8.9 μM), folfirinox (Oxa 37 μM, 5-FU 4.7 μM, Iri 6.2 μM), and gemcitabine (9.3 μM) for 10, 24, 48, and 58 hours. Data are normalized to *GAPDH* and are presented as fold change in gene expression relative to Ctrl. Statistical analysis of CTRL vs treated alginate tumor microgels, ** $p < 0.005$ and *** $p < 0.0005$.

**Conclusions**

In this work, we report a simple and reproducible cell microencapsulation technology to develop *in vitro* 3D alginate-based PDAC models for simultaneously monitoring metabolic pH variations and quantifying drug response. Optical ratiometric pH sensors, pancreatic tumor cells, and pancreatic stromal cells were embedded in alginate microgels and the extracellular pH fluctuations were monitored during chemotherapeutic treatment through real-time fluorescent quantitative microscopy. After live imaging, the quantification of the spatio-temporal pH gradients in the whole hydrogels during drug testing was obtained in a fast and non-invasive way by applying automated computational analyses. The anti-tumor effect of the PDAC standard therapies, such as gemcitabine, folfirinox, and paclitaxel, correlated with the extracellular acidification detected within the tumor-stroma microgels over time.

Because cell metabolic heterogeneity is crucial in tumor response to conventional drug therapies [75,76], this pH sensing platform is a powerful tool for understanding how drug sensitivity is influenced by the interplay between the different components of the microenvironment. Furthermore, the proposed platform allows to define the influence of the pH cellular metabolic variations on ~~the~~ drugs response widely used in PDAC therapy, taking into account the complex crosstalk existing within the tumor microenvironment. In particular, such fast correlation allows to predict the effectiveness of a drug, which could result in a remodulation of the existing chemotherapeutic agents or in new therapies for PDAC treatment. Interestingly, we propose the use of pH-sensing alginate microgels as an *in vitro* 3D tumor model for drug screening, personalized medicine, and for evaluating the acquisition of chemoresistance following drug treatment. Indeed, these systems can be integrated with patient-derived cells and employed to monitor metabolic changes during drug testing and predict the efficacy of anti-cancer drugs in a matter of hours *vs.* weeks in patients.

**Declaration of competing interest**
The authors declare no competing interests.

**ACKNOWLEDGEMENTS**
This work was supported by the European Research Council (ERC) under the European Union's Horizon 2020 research and innovation program ERC Starting Grant "INTERCELLMED" (contract number 759959), the European Union's Horizon 2020 research and innovation programme under grant agreement No. 953121 (FLAMIN-GO), the Associazione Italiana per la Ricerca contro il Cancro (AIRC) (MFAG-2019, contract number 22902; Bridge Grant AIRC 2022, contract number 27012), the "Tecnopolo per la medicina di precisione" (TecnoMed Puglia) - Regione Puglia: DGR n.2117 of 21/11/2018, CUP: B84I18000540002), the Italian Ministry of Research (MUR) in the framework of the National Recovery and Resilience Plan (NRRP), "NFFA-DI" Grant (B53C22004310006), "I-PHOQS" Grant (B53C22001750006) and under the complementary actions to the NRRP funded by NextGenerationEU, "Fit4MedRob" Grant (PNC0000007, B53C22006960001). The PRIN 2022 (2022CRFNCP_PE11_PRIN2022) funded by European Union – Next Generation EU, the Young Researchers-MSCA-PNRR-MUR (MSCA_0000023) and the Fondazione Umberto Veronesi. The authors gratefully thank the collaborator Paolo Cazzato (Institute of Nanotechnology, Lecce, Italy) for providing technical support.

**Appendix A. Supplementary data**
Supplementary data to this article can be found online at https://XXXXXX.